# Evidence Of Protein Collective Motions On The Picosecond Time Scale

*Yunfen He[1], Jing-Yin Chen[2], Joseph  R. Knab[3], Wenjun Zheng[1], Andrea G.  Markelz[1\*]*

[1]University at Buffalo, the State University of New York, [2]Institute for Shock Physics, Washington State University, [3]Delft University of Technology

1 **Abstract**

2 We investigate the presence of structural collective motions on a picosecond time scale for the

3 heme protein, cytochrome c, as a function of oxidation and hydration, using terahertz (THz)

4 time-domain spectroscopy and molecular dynamics simulations. The THz response dramatically

5 increases with oxidation, with the largest increase for lowest hydrations and highest frequencies.

6 For both oxidation states the THz response rapidly increases with hydration saturating above

7 ~25% (g $H_2O$/g protein).  Quasi-harmonic vibrational modes and dipole-dipole correlation

8 functions are calculated from molecular dynamics trajectories. The collective mode density of

9 states alone reproduces the measured hydration dependence providing strong evidence of the

10 existence of these motions. The large oxidation dependence is reproduced only by the dipole-

11 dipole correlation function, indicating the contrast arises from diffusive motions consistent with

12 structural changes occurring in the vicinity of a buried internal water molecule.

14 **Introduction**

15    Protein function relies on structural dynamics, with time scales ranging from picoseconds to

16 beyond seconds. The transitions to the different configurations involved in function are routinely

17 reproduced by trajectories involving only the first few structural collective vibrational modes,

18 suggesting the importance of these modes in understanding and tailoring of protein interactions.

19 However, there is some debate as to whether large scale structural collective motions exist and if

20 instead dynamics are entirely directed Brownian motion. Such uncorrelated diffusive motion

21 would need to serendipitously access the proper configuration for protein-protein or protein-

22 ligand binding. Concerted motions could explain the observed high physiological on-rates and

[Insert Running title of <72 characters]

1 affinities [1-4]. Recent neutron spin echo and X-ray inelastic scattering have verified that collective

2 motions do occur [5,6], however these challenging measurements did not address the influence of

3 these modes on function. The nature of these measurements limits their broad application for

4 protein engineering. The question of coupling of large scale motion is critical in the quest for

5 tailoring allosteric interactions.

6 Computational studies of collective motions include normal modes, quasiharmonic modes and

7 course grain modes. An intuitive example of large-scale collective motion is the hinging motion

8 of lysozyme. The hinge motion is immediately apparent as the upper and lower portions of the

9 protein clamp down upon the substrate resulting in more effective cleavage. However one can

10 also visualize the molecule without substrate and the hinge continuously oscillating. This motion

11 was first calculated over 20 years ago by Karplus and Brooks using normal mode analysis for

12 lysozyme, with a hinge bending resonant frequency of 3.6 cm$^{-1}$ = 0.108 THz [7]. While there is

13 some skepticism if these harmonic motions occur in vivo the calculated modes lead to root mean

14 squared displacements in good agreement with experimental results [8,9]. The calculated time

15 scales for these correlated motions can be from the ps to ns.

16 To fully characterize these motions requires an experimental tool which can both resolve

17 energy and momentum transfer in the picosecond range. Inelastic neutron scattering (INS) is

18 such a technique, however its application is limited since it is not a table top instrument and

19 requires large samples (~100 mg) and deuteration. Another relevant spectral technique is

20 terahertz time domain spectroscopy (THz TDS). Previously we and others have used terahertz

21 time domain spectroscopy to measure the dielectric function of protein samples. The dielectric

22 response has contributions from vibrational and diffusive motions in the protein and adjacent

23 solvent:

[Insert Running title of <72 characters]

1    $$\varepsilon(\omega) = \varepsilon_o + \int \frac{f(\omega')g(\omega')}{(\omega'^2 - \omega^2) + i\gamma(\omega')\omega} d\omega' + \varepsilon_r \int_0^\infty \frac{h(\tau)d\tau}{1 + i\omega\tau} \quad (1)$$

2    where $\varepsilon_o$ is the DC dielectric constant. The second term on the right hand side contains the

3    vibrational density of states (VDOS) $g(\omega)$, oscillator strength $f(\omega)$ and damping coefficient $\gamma(\omega)$.

4    The third term is the relaxational response, assuming Debye relaxation for a distribution of

5    relaxation times $h(\tau)$. Typically, what is measured is the absorption coefficient $\alpha(\omega)$ and the

6    refractive index $n(\omega)$. Relating these measured quantities to the vibrational and relaxational

7    response, one obtains:

9    $$\alpha_{vib}(\omega) = \int \frac{g(\omega')f(\omega')\omega\gamma(\omega')}{(\omega'^2 - \omega^2)^2 + \gamma(\omega')^2\omega^2} d\omega' \quad \text{and} \quad \alpha_{relax}(\omega)n_{relax}(\omega) = \int \frac{1}{c} \frac{h(\tau)\tau\omega^2}{1 - \tau^2\omega^2} d\tau \quad (2)$$

11    where $\alpha_{vib}(\omega)$ is the absorption coefficient which arises from the vibrational motions and

12    $\alpha_{relax}(\omega)n_{relax}(\omega)$ is the product of the absorption coefficient and refractive index arising from the

13    relaxational response. The relaxational response is always a broad absorbance. For small

14    systems (5-20 atoms), the frequency dependence of the vibrational and relaxational responses are

15    distinct, with the vibrational resonances widely spaced and relatively narrow. For large

16    macromolecules, the collective vibrational modes become dense and individual modes begin to

17    overlap. The spectrum becomes sufficiently dense that it is difficult to distinguish the structural

18    vibrational response from localized side chain relaxational rotations[10]. Such a distinction is

19    important for applications such as allosteric inhibitor design, where the binding of the inhibitor is

20    constructed to interfere with the structural motions necessary for function [11]. Our ability to

21    characterize global structural motions distinct from local diffusive ones is challenged by the

[Insert Running title of <72 characters]

1   similarity in the relaxational and vibrational response for large macromolecules. However the

2   manner in which these different types of motions change with environment or functional state

3   can be distinct. To test for the presence and contribution of picosecond collective vibrational

4   modes, we compare systematic measurements of cytochrome c (CytC) as a function of hydration

5   and oxidation with the calculated response from the collective vibrational motions and the

6   dipole-dipole correlation function which includes relaxational response. We find collective

7   vibrational motions account for the hydration dependence observed in the THz dielectric

8   response. However, the frequency dependence appears to be dominated by the relaxational

9   motions of water and side chains. Furthermore the large oxidation dependent dielectric response

10   observed experimentally is only reproduced when relaxational motions are included.

11      CytC is a protein of increasing interest as its multiple functions become revealed. The

12   primary role of CytC is to participate in the metabolism process in the mitchondria through the

13   transfer of an electron from cytochrome c reductase to cytochrome c oxidase, both embedded in

14   the inner mitochondrial membrane. Recently, it has become clear that CytC plays an important

15   role in apoptosis, both within the mitchondria and during its release into the cytosol [12].

16   Furthermore, CytC is an excellent model heme protein to consider fundamental questions of

17   protein dynamics, as it is sufficiently small for systematic comparisons to theoretical modeling.

18      The physiochemical properties of CytC can be explained in terms of the differences in the

19   dynamic behavior of the two redox states. Eden and co-workers proposed that the oxidized form

20   of CytC is more flexible than the reduced form [13], based on an estimated 40% increase in the

21   apparent compressibility of CytC upon oxidation. X-ray diffraction measurements appeared to

22   give some support to higher flexibility in the ferri state, as the atomic mean squared displacement

23   (msd) as measured by the Debye Waller factor increases for ferri over ferro [14]. The motions

[Insert Running title of <72 characters]

1    contributing to msd can be both collective as in structural vibrational modes and local as in side

2    chain librations.  A quantitative relationship can be drawn between the collective modes and the

3    structural flexibility using the frequency dependence of the vibrational density of states (VDOS).

4    Low frequency modes below $k_BT$ are thermally occupied and these motions will contribute to the

5    msd, whereas high frequency modes with energies on the order of $k_BT$ or higher have lower

6    occupation and have a smaller contribution to the msd.  THz-TDS was previously used to

7    investigate if the implied changes in the VDOS occurs with the CytC flexibility change with

8    oxidation [15].  A large increase in the THz dielectric response was observed with oxidation

9    consistent with an increase in the low frequency VDOS and a higher flexibility.  A number of

10   observations since those studies have suggested caution when directly relating the dielectric

11   response to the VDOS.  These include the exploration of the role of local relaxational motions in

12   the THz dielectric response in hydration measurements on lysozyme [16];  nuclear vibrational

13   resonance spectroscopy (NRVS) measurements showing a very slight increase in the VDOS for

14   the modes coupled to the heme Fe with oxidation of CytC [17]; and the report of a large

15   enhancement in the dielectric response for water immediately adjacent to the protein over that of

16   bulk water [18].  This last point is critical in that if the equilibrium water content is dependent on

17   oxidation state, the THz contrast observed may arise from the different water contributions and

18   not from a density of states change.  It is therefore important that the comparison between

19   oxidation states be made at equivalent water content.

20       More importantly the systematic hydration and oxidation dependence data allows us to test the

21   degree structural vibrational modes contribute to the terahertz response and to determine the

22   origin of the oxidation dependence by comparison with molecular dynamics (MD) simulations.

23   We present two simulation approaches, both using calculated molecular trajectories:  harmonic

[Insert Running title of <72 characters]

1 vibrational response based on quasiharmonic analysis; and the full response as determined by the

2 power spectrum of the dipole-dipole correlation function. Quasiharmonic analysis is also

3 referred to as principle component analysis (PCA). PCA is essentially a harmonic analysis

4 allowing for changes in the effective force constants at finite temperature. In this method, the

5 MD simulation is utilized to obtain effective modes of vibration from the atomic fluctuations

6 about an average structure. These modes include the anharmonic effects neglected in a normal

7 mode calculation [19]. PCA introduces temperature dependent anharmonicity, but still calculates

8 the harmonic vibrational modes of the system as well as the dipole derivative for each mode,

9 which is used to calculate the absorption coefficient. This analysis does not capture diffusive

10 motion such as librational motion of side chains and individual rotational motion of solvent

11 molecules. To capture the full dielectric response involving all motions one must use the

12 complete trajectory calculated from the full potential. The dipole-dipole correlation function is

13 calculated from the MD trajectories and then the Fourier transform of the correlation function

14 reveals the frequency dependent absorption coefficient. We perform both types of calculations

15 as a function of hydration and oxidation state and compare with the measured results.

16  We find the measured hydration dependence of the picosecond dynamics for both oxidation

17 states of CytC deviates from the typical rapid increase at 30 % wt, and that the VDOS calculated

18 using PCA reproduces the observed hydration dependence. This ability to reproduce the

19 measured hydration dependence with the quasiharmonic modes indicates the presence of

20 collective modes at THz frequencies. However PCA does not reproduce the observed oxidation

21 or frequency dependence. We find that the calculated dielectric response from the full dipole-

22 dipole correlation function which includes diffusive motions does reproduce the oxidation and

23 frequency dependence. We propose that the THz dielectric response is dominated by the

[Insert Running title of <72 characters]

1  anharmonic motions such as surface side chain rotational motions or possibly motion of an

2  internally bound water molecule.

4  **Experimental Details**

6  Sample Preparation and Characterization:

7  Preparation of oxidized and reduced CytC films measured in THz time domain spectroscopy is

8  described in a previous study[15]. Solutions were pipetted onto clean infrasil quartz substrates with

9  half the substrate left bare for referencing. The films were characterized by UV/Vis absorption

10  to verify the oxidation state (see supplementary material). The thickness of the dried films is

11  measured to both verify uniformity and to determine absolute absorption coefficients and

12  refractive indices. Film thicknesses were typically on the order of 100 m with thickness

13  variation < 5% for a given film. See ref [20] for method of film thickness determination.

15  Isotherm Determination:

16  To compare the picosecond dielectric response of ferro and ferri samples at equivalent water

17  content we measured the isotherms for both oxidation states. The water content for a given

18  relative humidity often follows the B.E.T. thermal equation[21]:

19
$$h(x) = \frac{ah_m x}{(1+x)[1+(a-1)x]} \qquad (3)$$

20  where $h(x)$ is the water content as % wt, that is, g water/100 g protein; $a$ is a parameter

21  representing the potential water absorption capacity of the protein; $h_m$ is a parameter representing

22  the humidity for monolayer of water; and x is the relative pressure of the gas, namely, the

23  relative humidity. See supplementary material for method and results.

[Insert Running title of <72 characters]

2      THz TDS:

3      An $N_2$ purged terahertz time domain spectroscopy system (THz TDS) with bandwidth 5 cm$^{-1}$ -

4      86 cm$^{-1}$ is used to monitor the change in absorbance and index with oxidation states of CytC

5      films at different hydration.  The THz radiation is generated by a Hertzian dipole antenna and

6      detected electro-optically.[22,23]  All measurements were performed at room temperature and under

7      hydration control.  Substrates were mounted on a brass holder with two apertures for the

8      reference bare substrate and the CytC film.  The plate is mounted in a closed hydration cell with

9      the relative humidity controlled by flushing the cell with a hydrated gas from a Licor dewpoint

10     generator.  Nitrogen (air) was used as the flushing gas for the ferro (ferri) films.  The closed

11     hydration cell is placed at the focus of the THz TDS system.

12     We measure the terahertz dielectric response of both oxidized and reduced CytC films as a

13     function of hydration.  The time required for CytC films to reach hydration equilibrium is

14     determined by TGA (thermogravimetric analysis) and terahertz transmission and both found the

15     time for.dehydration was ~ ½ hour and for hydration ~ 1 hour.  To ensure equilibrium hydration

16     the samples were exposed to a given relative humidity in the hydration cell for at least one hour

17     for film dehydration and two hours for film hydration.  The vertically mounted substrates limited

18     the highest hydration to below the film's threshold for flowing.  For ferri (ferro) -CytC films, the

19     highest water content achieved was 75 % wt (57 % wt).  A transmission measurement consists of

20     toggling between the sample and reference apertures.  Absorption due to the gas phase water of

21     the humidity cell is removed by the reference which is in the same humidity cell as the sample.

22     The real part of the refractive index and the absorption coefficient are extracted from the

23     terahertz transmission data in the standard way [24].

       [Insert Running title of <72 characters]

2    Molecular Dynamics Calculations:

3    We used Version 32 of CHARMM[25] with all-atom parameter set 22 [26].  The x-ray structure file

4    1hrc.pdb was used as the starting structure.  Heme group partial charges for ferro and ferri states

5    are taken from reference [27]. The heme group is patched to CytC though axial ligated bounds, Met

6    80 and His 18 and covalent bounds, Cys14 and Cys17. Different hydrations are simulated by

7    solvating the protein with a layer of water molecules with different thicknesses between 1.5 Å

8    and 7 Å by using the "SOAK" command in Insight II. The structures of the solvated proteins are

9    built in CHARMM by using TIP3 force field. The energy minimized structure was then used for

10   MD simulations.

11   MD simulations were carried out with an integration time step of 0.001 ps. The system was

12   heated to 300 K with a temperature increment of 100 K after each 1000 steps. Temperature

13   equilibration was followed by a constant temperature MD run. The trajectory is 1ns, starting with

14   399 ps equilibrium run, being followed by a 601ps production run.  We find the water rapidly

15   moves to equilibrium positions with the water structure resembling X-ray measurements within

16   50 ps.

17   From the production run of the MD simulation the average structure of each protein was

18   determined separately using the corresponding 6001 frames and, superimposing each structure

19   onto the average structure, a quasiharmonic vibrational analysis was performed. This is done by

20   diagonalizing the whole covariance matrix for a given protein.  The calculation output includes

21   the eigenvalues (mode frequency), the eigenvectors, and the dipole derivatives for each mode [28].

22   For mode frequency $\nu_k$, the integrated intensity $\Gamma_k$ has units of molar absorptivity and is

23   proportional to the absorption coefficient [29].  We draw a distinction of $\Gamma_k$ with the often

[Insert Running title of <72 characters]

1 calculated absorption intensity, $A_k$. $A_k$ differs from $\Gamma_k$ by a factor of $\nu_k$ and is not directly related

2 to the derivation of the absorption coefficient from Fermi's Golden rule. The double harmonic

3 approximation consists of modes calculated from a harmonic expansion of the potential and an

4 approximation of the dipole moment inner product using the quadratic term in the Taylor

5 expansion of the dipole.  can be obtained from the dipole derivatives calculated from the

6 quasiharmonic analysis using[30]:

7
$$\Gamma_k = \frac{N_0 \pi^2}{3c^2 \varepsilon_0 \omega_k} \left( \frac{\partial p}{\partial Q_k} \right)^2 \qquad (4)$$

8 where $\varepsilon_0$ is the permittivity of vacuum, and $N_0$ is the Avogadro number and the magnitude of

9 the dipole derivative

10 The net calculated absorption coefficient is proportional to the sum of Lorentzian oscillators

11 with relaxation rates  such that we define our quasi harmonic calculated absorption coefficient

12 as

13
$$\alpha_{QH}(\omega) = \sum_k \frac{1}{\pi} \frac{\Gamma_k \gamma^2}{(\omega - \omega_k)^2 + \gamma^2}. \qquad (5)$$

14 $\gamma$ is is set to $4\,cm^{-1}$ in agreement with the FWHM observed for THz vibrational resonances

15 for molecular crystals.

16 The absorption coefficient per unit length $\alpha(\omega)$ and the refractive index $n(\omega)$ are related to the

17 imaginary part of the dielectric constant $\varepsilon''(\omega)$ by $\alpha(\omega)n(\omega) = (\omega/c)\,\varepsilon''(\omega)$. Within linear-

18 response theory, $\alpha(\omega)\,\nu(\omega)$ is given by the power spectrum of the time-correlation function of

19 the total dipole operator [31]:

20
$$\alpha(\omega)n(\omega) = \frac{2\pi\omega^2\beta}{3cV} \int_{-\infty}^{\infty} dt e^{-i\omega t} < M(0)M(t) > \qquad (6)$$

[Insert Running title of <72 characters]

1    Where *M* is the total dipole moment of the system, *V* is the volume of the system, and where

2    $\beta = (k_BT)^{-1}$ is the inverse temperature. We calculate the time dependence of the total dipole

3    moment *M(t)* for the trajectory assuming constant partial charges for all atoms during the entire

4    trajectory.

6    **Results**

7    In Figure 1, we show the measured frequency dependence of the THz absorption coefficient.

8    The THz absorption of the ferri-CytC and ferro-CytC films increases monotonically with

9    frequency.    There are no distinct resonances observed in    , however there is an obvious

10   increase with oxidation. The refractive index of CytC films is nearly frequency independent up

11   to 85 cm$^{-1}$ and also has a large increase with oxidation (n$_{ferro}$ = 1.65 to n$_{ferri}$ = 1.80 for 3% h).

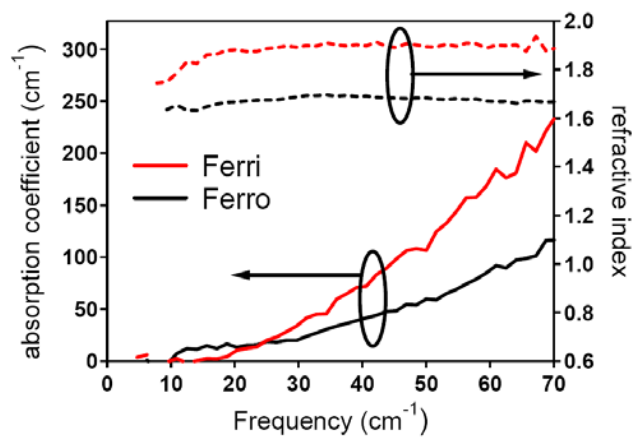

13   Fig. 1.

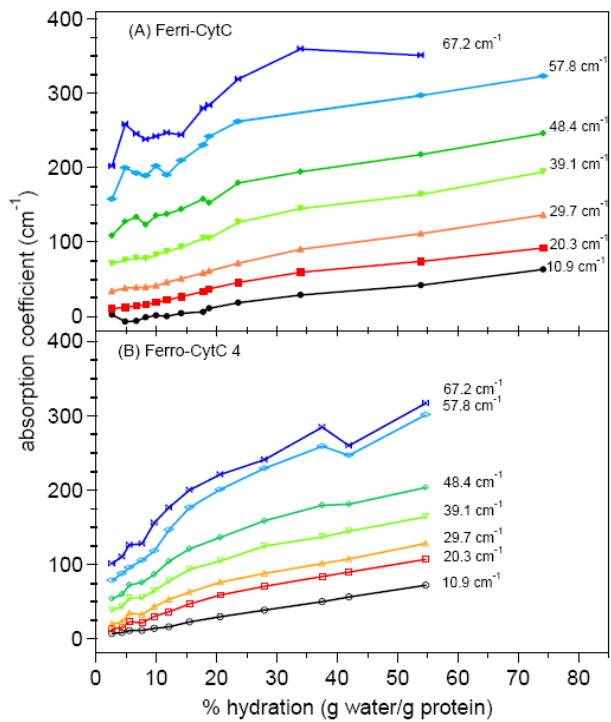

2    Fig. 2

3    We compare the absorption coefficients of ferri-CytC and ferro-CytC films as a function of

4    water content in Figure 2.  The absorption coefficient rapidly increases with hydration with the

5    rate of increase abruptly decreasing at a cross over point ~ 20-25% wt.  This cross over behavior

6    from a rapid increase with hydration to a saturation behavior for both oxidation states is readily

7    apparent in the refractive index data in Figure 3.

8    Comparing the response at equivalent water content we see in Figure 2 and 3, an obvious

9    oxidation dependence of the picosecond response as was reported earlier.  The dependence is

10   most dramatic for the index measurements, where the ferri state values are consistently larger for

11   all hydrations and frequencies.  While the absorption coefficient does increase in the ferri state at

12   higher frequencies and lower hydrations, this contrast decreases with increasing hydration and at

13   lower frequencies.  This oxidation dependence is consistent with Takano and Dickerson's B-

[Insert Running title of <72 characters]

1 factors where the average B-factor was found to increase from 18.7 in the ferro state, to 22.7 in

2 the ferri state [14].

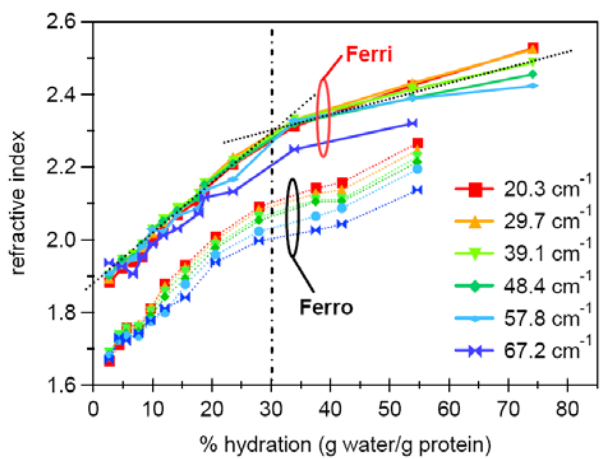

4 Fig. 3

5      We compare the measured response to the calculated response for only collective modes and

6 for the full response including local relaxational motions. In addition to PCA calculations

7 presented here, we performed the simpler normal mode analysis where the force constants are

8 extracted from the curvature of the potential at the minimized energy. We found little agreement

9 between the calculated normal modes and the measured frequency, hydration or oxidation

10 dependence indicating that normal mode analysis is inadequate for accurately describing the

11 dynamics at room temperature.

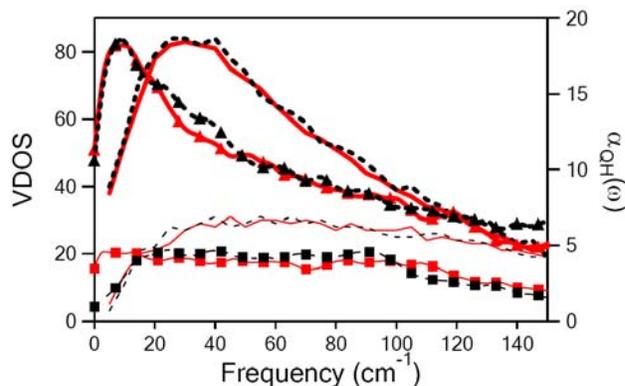

Fig. 4

We first examine the frequency and oxidation dependence calculated using PCA. The vibrational density of states (VDOS) and absorptivity for oxidized and reduced CytC are shown in Figure 4. The VDOS increases with frequency reaching a peak at ~ 30 cm$^{-1}$ and then decreases. As the hydration increases, the number of quasiharmonic modes for oxidized and reduced CytC also increases and the peak in the VDOS red shifts. This frequency dependence does not resemble the measured dielectric response seen in Figure 1. Similarly there is no apparent oxidation dependence in the VDOS. The absorptivity is calculated from the dipole derivative of the mode. It is assumed the partial charges of the individual atoms remain constant as the atoms move along the eigenvector, which is a problematic but common assumption. Again neither the frequency or oxidation dependence of the calculated absorptivity from the collective modes resembles the measured response.

[Insert Running title of <72 characters]

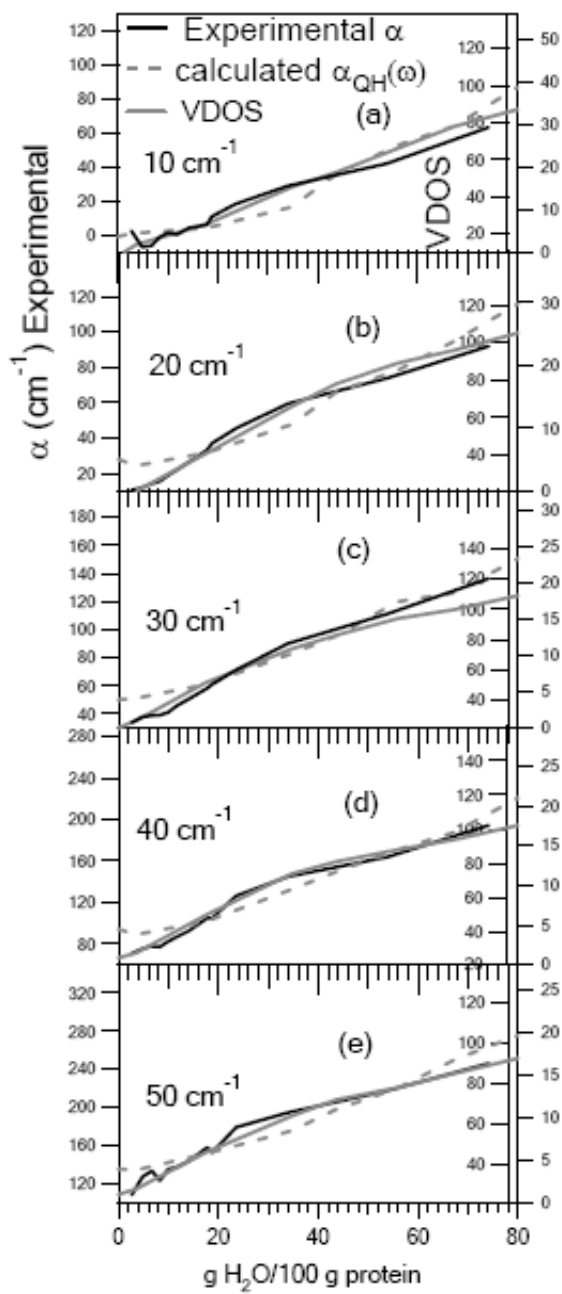 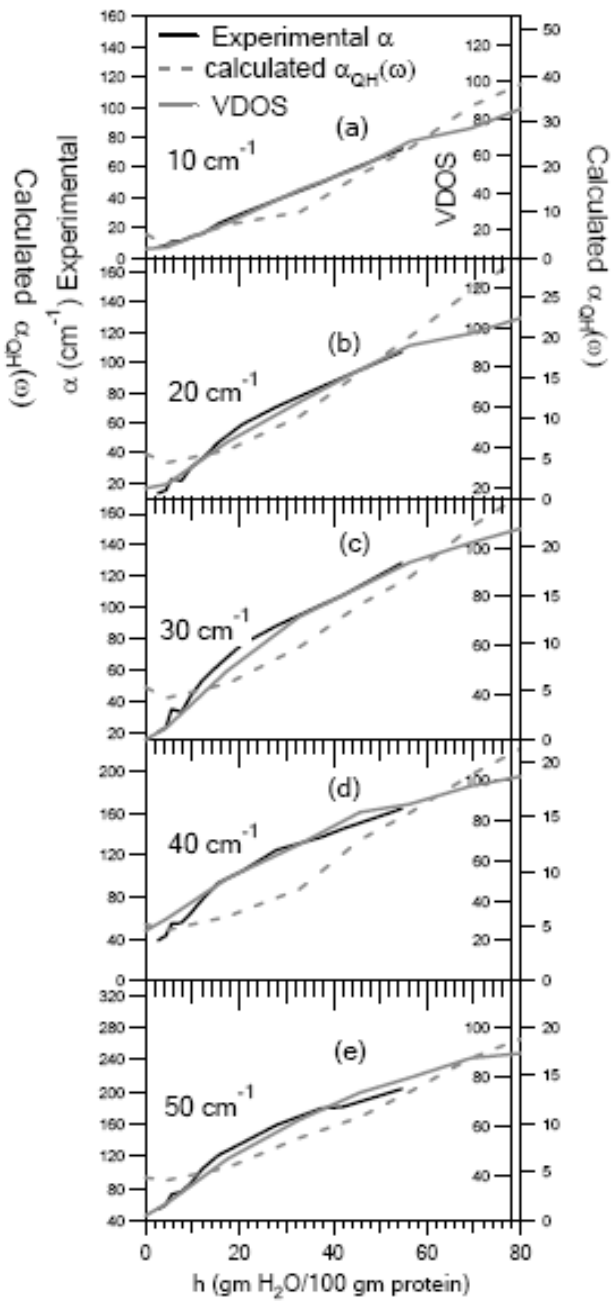

Fig. 5                                         Fig. 6

Figures 5 (6) shows a comparison of the hydration dependence of the measured absorption coefficient, the calculated VDOS and the calculated absorption coefficient for several representative frequencies for ferri (ferro) CytC.  The observed hydration dependence for cytochrome c departs from the expected hydration dependence of protein dielectric response as

[Insert Running title of <72 characters]

1    characterized over a broad frequency range up to 10 GHz 40 years ago [32]. There, for the proteins

2    studied, the dielectric response increases slightly with hydration up to 30% wt then increases

3    rapidly above this hydration.  The rapid increase at 30 % wt is associated with anharmonic

4    motions accessible with the increased plasticity with hydration [33-36].  This departure for the

5    hydration dependence of CytC response provides a test for the MD simulations. We see in Figs.

6    5 and 6 the collective mode VDOS reproduces the hydration dependence well for both oxidation

7    states and over the entire frequency range.  This agreement suggests A) the presence of structural

8    vibrational modes contributing to the dielectric signal and B) these modes are responsible for the

9    observed hydration dependence.   The hydration dependence of the calculated absorption

10   coefficient does not reproduce the measurements as well.  It is possible that this is in part due to

11   the assumption of constant partial charges in the dipole derivative calculations.

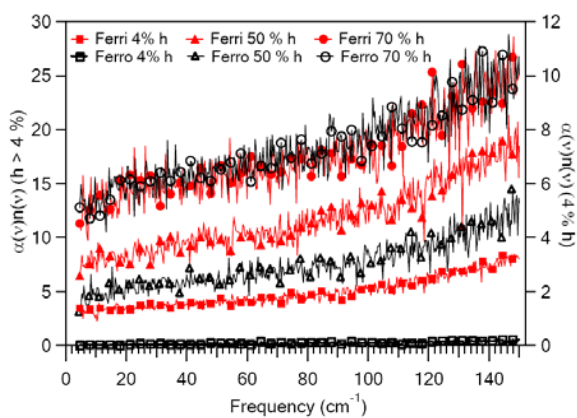

13   Fig. 7

14   It is clear that the collective modes do not give rise to the oxidation sensitivity of the THz

15   response.  The PCA calculations for this frequency range account only extended oscillatory

16   motions.   The Fourier transform of the dipole-dipole correlation gives the product of the

17   frequency dependent absorption coefficient and the refractive index, $\alpha(\omega)v(\omega)$ which includes all

18   motions contributing to the response, including local relaxational motions of water and surface

[Insert Running title of <72 characters]

1  side chains.  As shown earlier, the measured refractive index has little frequency dependence so

2  we directly compare the calculated $\alpha(\omega)v(\omega)$ with the measured $\alpha(\omega)$.  In Figure 7 we show the

3  frequency dependence for $\alpha(\omega)v(\omega)$ several hydrations.  The frequency dependence now

4  resembles the measured absorption coefficient and we see that at low hydration there is a strong

5  contrast between oxidized and reduced cytochrome c.  As the hydration increases this difference

6  decreases.  At the highest hydration levels one cannot distinguish between the two oxidation

7  states.

9  **Discussion**

10  The hydration dependent collective mode calculations indicate their contribution to the

11  dielectric response.  As seen in Figure 4, the collective modes do not reproduce the frequency

12  and oxidation dependencies.  While the modes at increasingly higher frequencies progressively

13  more localized, the motions even up to 300 cm$^{-1}$ are extended beyond individual residues (see

14  supplemental material).  The quasiharmonic VDOS never captures the low frequency diffusive

15  motions of surface side chains.  The residence times for these motions were calculated with

16  typical times ~ 2-10 ps [37]. These motions have been discussed by Sokolov and others when

17  analyzing the anharmonicity in the neutron temperature dependence [38] and will be present in the

18  calculated full trajectory.  That the observed oxidation state dependence is not reproduced by any

19  of the harmonic analysis, but is reproduced by the dipole-dipole correlation indicates that the

20  oxidation dependence arises from relaxational motions.  We suggest that the internal bound

21  water dynamics are influenced by the local electrostatics and it is these motions that give rise to

22  the oxidation dependence seen in the THz response.  At low hydrations these few internally

23  bound waters provide a large contribution to the overall signal and therefore an oxidation

[Insert Running title of <72 characters]

1  dependence is observed.  At higher hydrations the surface water, mobile surface side chains and

2  the vibrational modes begin to dominate the signal, and the internal water has little net

3  contribution, so the oxidation dependence is no longer significant at the higher hydrations.

4      We note the contrast seen in the THz measurements is not seen by another important protein

5  dynamics characterization technique nuclear resonant vibrational spectroscopy (NRVS) [39].  An

6  extensive NRVS study found only a very slight increase with oxidation for CytC [17,40].

7  Accompanying this, MSD simulations for the heme group and backbone had no discernible

8  oxidation dependence.  These calculations did not include the diffusive water or side chain

9  motion. The oxidation dependence arising from internal water diffusive motion could explain

10  why little oxidation dependence is seen by NRVS, as there the samples were fully hydrated and

11  the motions observed are vibrational modes coupled to the heme Fe.

13  **Conclusion**

14      To conclude, terahertz spectroscopy is sensitive to protein environment and functional state.

15  Structural collective modes as calculated by PCA reproduce the measured hydration dependence

16  of absorption coefficient, demonstrating their contribution to the picosecond response.  The

17  oxidation dependence of the picosecond response is only reproduced by analysis of the complete

18  trajectory, demonstrating that it is manifest only in the diffusive motions of either the side chains

19  or the internal water.  The evidence of structural collective modes in these smaller molecular

20  weight proteins is consistent with recent results using neutron spin echo measurements (NSE) on

21  somewhat more massive proteins [6,41].  The terahertz spectroscopy approach may provide a

22  critical complementary technique to NSE for collective modes on shorter time scales.

[Insert Running title of <72 characters]

**Acknowledgements**

This work was supported by ARO grant DAAD 19-02-1-0271, ACS grant PRF 39554-AC6, NSF CAREER grant PHY-0349256, NSF REU grant DMR-0243833 and NSF IGERT grant DGE0114330.

**References**

(1)     Goodey, N. M.; Benkovic, S. J. *Nature Chemical Biology* **2008**, *4*, 474-82.

(2)     Hammes-Schiffer, S.; Benkovic, S. J. *Annu. Rev. Biochem.* **2006**, *75*, 519-41.

(3)     Lange, O. F.; Lakomek, N.-A.; Farès, C.; Schröder, G. F.; Walter, K. F. A.; Becker, S.; Meiler, J.; Grubmüller, H.; Griesinger, C.; Groot, B. L. d. *Science* **2008**, *340*, 1471-75.

(4)     Jarymowycz, V. A.; Stone, M. J. *Chem. Rev.* **2006**, *106*, 1624-1671.

(5)     Liu, D.; Chu, X.-q.; Lagi, M.; Zhang, Y.; Fratini, E.; Baglioni, P.; Alatas, A.; Said, A.; Alp, E.; Chen, S.-H. *Phys. Rev. Lett.* **2008**, *101*, 135501.

(6)     Biehl, R.; Hoffmann, B.; Monkenbusch, M.; Falus, P.; Pre´ost, S.; Merkel, R.; Richter, D. *Phys. Rev. Lett.* **2008**, *101*, 138102.

(7)     Brooks, B.; Karplus, M. *Proceedings of the National Academy of Sciences of the United States of America* **1985**, *82*, 4995-4999.

(8)     Teeter, M. M.; Case, D. A. *J. Phys. Chem.* **1990**, *94*, 8091-97.

(9)     Zoete, V.; Michielin, O.; Karplus, M. *J. Mol. Biol.* **2002**, *315*, 21-52.

[Insert Running title of <72 characters]

1    (10)    Markelz, A. G. *IEEE J. Sel. Topics in Quantum Electronics* **2008**, *14*, 180-190.

2    (11)    Kern, D.; Zuiderweg, E. R. *Current Opinion in Structural Biology* **2003**, *13*, 748–
3    757.

4    (12)    Kagan, V. E.; Bayir, A.; Bayir, H.; Stoyanovsky, D.; Borisenko, G. G.; Tyurina,
5    Y. Y.; Wipf, P.; Atkinson, J.; Greenberger, J. S.; Chapkin, R. S.; Belikova, N. A. *Mol. Nutr.*
6    *Food Res.* **2009**, *53*, 104-114.

7    (13)    Eden, D. M., James B.; Rosa, Joseph J.; Richards, Frederic M. *Proceedings of the*
8    *National Academy of Sciences of the United States of America* **1982**, *79*, 815-819.

9    (14)    Takano, T.; Dickerson, R. E. *Proceedings of the National Academy of Sciences of*
10   *the United States of America* **1980**, *77*, 6371-6375.

11   (15)    Chen, J.-Y.; Knab, J. R.; Cerne, J.; Markelz, A. G. *Phys. Rev. E. Rapid* **2005**, *72*,
12   040901.

13   (16)    Knab, J. R.; Chen, J. Y.; He, Y.; Markelz, A. G. *Proc. of the IEEE* **2007**, *95*,
14   1605-10.

15   (17)    Leu, B. M.; Ching, T. H.; Zhao, J.; Sturhahn, W.; Alp, E. E.; Sage, J. T. *J. Phys.*
16   *Chem. B* **2009**, *113*, 2193–2200.

17   (18)    Ebbinghaus, S.; Kim, S. J.; Heyden, M.; Yu, X.; Heugen, U.; Gruebele, M.;
18   Leitner, D. M.; Havenith, M. *Proc. Natl. Acad Sci. U.S.A.* **2007**, *104*, 20749–20752.

19   (19)    Balog, E.; Becker, T.; Oettl, M.; Lechner, R.; Daniel, R.; Finney, J.; Smith, J. C.
20   *Phys. Rev. Lett.* **2004**, *93*, 28103.

[Insert Running title of <72 characters]

(20)     Whitmire, S. E.; Wolpert, D.; Markelz, A. G.; Hillebrecht, J. R.; Galan, J.; Birge, R. R. *Biophys. J.* **2003**, *85*, 1269-1277.

(21)     Gascoyne, P. R. C.; Pethig, R. *J. Chem. Soc. Faraday Trans.* **1977**, *73*, 171-180.

(22)     Grischkowsky, D.; Katzenellenbogen, N. In *OSA Proceedings on picosecond electronics and optoelectronics*; Sollner, T. C. L., Shah, J., Eds.; OSA: Washington, DC, 1991; Vol. 9.

(23)     Jiang, Z.; Zhang, X.-C. *IEEE TRANSACTIONS ON MICROWAVE THEORY AND TECHNIQUES* **1999**, *47*, 2644-2650.

(24)     Knab, J.; Chen, J.-Y.; Markelz, A. *Biophys. J.* **2006**, *90*, 2576-2581.

(25)     Brooks, B. R.; Bruccoleri, R. E.; Olafson, B. D.; States, D. J.; Swaminathan, S.; Karplus, M. *J. Comput. Chem.* **1983**, *4*, 187-217.

(26)     MacKerell, A. D., Jr.; Bashford, D.; Bellott, M.; Dunbrack, R. L.; Evanseck, J. D.; Field, M. J.; Fischer, S.; Gao, J.; Guo, H.; Ha, S.; Joseph-McCarthy, D.; Kuchnir, L.; Kuczera, K.; Lau, F. T. K.; Mattos, C.; Michnick, S.; Ngo, T.; Nguyen, D. T.; Prodhom, B.; Reiher, W. E., III; Roux, B.; Schlenkrich, M.; Smith, J. C.; Stote, R.; Straub, J.; Watanabe, M.; Wiorkiewicz-Kuczera, J.; Yin, D.; Karplus, M. *J. Phys. Chem. B* **1998**, *102*, 3586-3616.

(27)     Autenrieth, F.; Tajkhorshid, E.; Baudry, J.; Luthey-Schulten, Z. *Journal of Computational Chemistry* **2004**, *25*, 1613-1622.

(28)     Levy, R. M.; Rojas, O. d. l. L.; Friesner, R. A. *J. Phys. Chem.* **1984**, *88*, 4233-4238.

[Insert Running title of <72 characters]

1    (29)    Person, W. B.; Zerbi, G. *Vibrational Intensities in Infrared and Raman*
2    *Spectroscopy*; Elsevier Scientific Publishing, 1982.

3    (30)    Galabov, B. S.; Dudev, T. *Vibrational Intensities*; Elsevier Science: Amsterdam,
4    1996.

5    (31)    McQuarrie, D. A.; First ed.; University Science Books: Sausalito, CA, 2000.

6    (32)    Harvey, S. C.; Hoeskstra, P. *J. Phys. Chem.* **1972**, *76*, 2987-2994.

7    (33)    Pethig, R. *Dielectric and electronic properties of biological materials*; First ed.;
8    Wiley: New York, 1979.

9    (34)    Bone, S.; Pethig, R. *Journal of Molecular Biology* **1982**, *157*, 571-575.

10    (35)    Bone, S.; Pethig, R. *Journal of Molecular Biology* **1985**, *181*, 323-326.

11    (36)    Rupley, J. A.; Careri, G. *Adv. Protein Chem.* **1991**, *41*, 37-172.

12    (37)    Best, R. B.; Clarke, J.; Karplus, M. *J. Mol. Biol.* **2005**, *349*, 185-203.

13    (38)    Roh, J. H.; Novikov, V. N.; Gregory, R. B.; Curtis, J. E.; Chowdhuri, Z.; Sokolov,
14    A. P. *Phys. Rev. Lett.* **2005**, *95*, 038101.

15    (39)    Achterhold, K.; Keppler, C.; Ostermann, A.; Bu¨rck, U. v.; Sturhahn, W.; Alp, E.
16    E.; Parak, F. G. *Phys. Rev. E* **2002**, *65*, 051916.

17    (40)    Leu, B. M.; Zhang, Y.; Bu, L.; Straub, J. E.; Zhao, J.; Sturhahn, W.; Alp, E. E.;
18    Sage, J. T. *Biophysical Journal* **2008**, *95*, 5874–5889.

[Insert Running title of <72 characters]

1   (41)   Bu, Z.; Biehl, R.; Monkenbusch, M.; Richter, D.; Callaway, D. J. E. *Proc Natl*

2   *Acad Sci U S A.* **2005**, *102*, 17646–17651.

3
4

5   **Legends**

6   Figure 1. Frequency dependence of THz absorption coefficient and index. The THz absorption

7   coefficient and index are shown for Ferri-CytC film and a Ferro-CytC film at different

8   hydrations as a function of frequency.

9   Figure 2. Hydration dependence of THz absorption coefficients. THz absorption coefficients of

10  Ferri (A) and Ferro (B) CytC are shown for several representative frequencies as a function %

11  hydration.  The data shows a rapid increase and then saturation of the response with hydration.

12  Figure 3.  Hydration dependence of THz refractive indices. Refractive indices of Ferri (A) and

13  Ferro (B) CytC are shown for several representative frequencies as a function of % hydration.

14  Lines drawn as guide to the eye.  The rapid turn over of the hydration dependence is nearly

15  frequency independent in contrast to the absorption coefficients shown in Figure 2.

16  Figure 4.  Frequency and oxidation dependence of the calculated collective modes.  The VDOS

17  and absorbtivity calculated from PCA is shown as a function of frequency for ferri (red solid

18  lines) and ferro (black dashed lines).  The lines without markers are for the VDOS and the lines

19  with markers are for absorptivity.  Thick lines are for 50% h and thin lines for 4% h.  Neither the

20  oxidation dependence nor the frequency dependence are the same as that observed.

[Insert Running title of <72 characters]

1  Figure 5. Comparison of the calculated structural collective mode to the measured hydration

2  dependence.   The measured absorption coefficient, calculated quasiharmonic VDOS, and

3  calculated absorbance $\alpha_{QH}(\omega)$ for Ferri-CytC is shown as a function of hydration for several

4  representative frequencies.   The calculated VDOS has good agreement with the measured

5  hydration dependence, suggesting the hydration dependence arises from these motions.

7  Figure 6. Comparison of the calculated structural collective mode to the measured hydration

8  dependence.   The measured absorption coefficient, calculated quasiharmonic VDOS, and

9  calculated absorbance $\alpha_{QH}(\omega)$ for Ferro-CytC is shown as a function of hydration for several

10  representative frequencies.   The calculated VDOS agrees well with the experimental results

11  similar to the ferri results in Figure 4. The agreement suggests the hydration dependence arises

12  from these motions.

14  Figure 7. The frequency dependent $\alpha(\omega)n(\omega)$ calculated from dipole-dipole correlation for

15  oxidized CytC (red, filled symbols) and reduced CytC (black, empty symbols) for several

16  hydration levels.   The right axis corresponds to h < 4 %.   The left axis corresponds to the

17  hydration >46%.  The 76 % h ferri and 78 % h ferro data are both offset by the same 5 units to

18  avoid overlap with the lower hydration data.   At the lower hydration levels the ferro CytC

19  response is less than the ferri CytC.   As the hydration increases, the difference with oxidation

20  decreases.   The calculated response includes both local diffusive motions and long range

21  collective motions.

[Insert Running title of <72 characters]

1   **Table of Contents Figure**

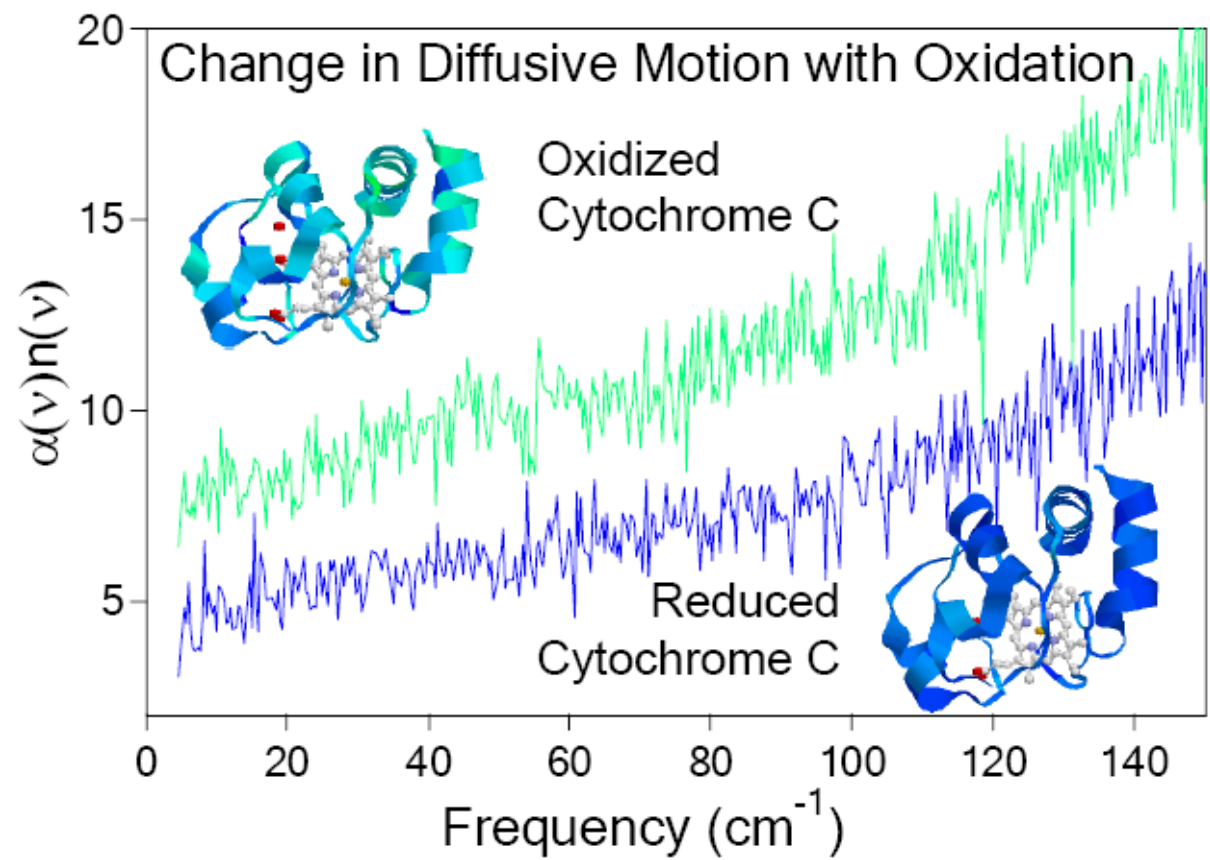

[Insert Running title of <72 characters]